\RequirePackage{amsmath}
\documentclass[runningheads]{llncs}
\usepackage{graphicx}
\usepackage{hyperref}

\usepackage{bbding}

\usepackage{amsmath}
\usepackage{amssymb}
\usepackage{mathtools}
\usepackage{booktabs}
\usepackage[binary-units = true, free-standing-units=true]{siunitx}
\DeclareSIUnit\px{px}

\DeclarePairedDelimiter\abs{\lvert}{\rvert}%

\usepackage{soul,xcolor}
 \newcommand{\eva}[1]{{\color{black} #1}}

\begin{document}
\title{\eva{3D Segmentation Networks for Excessive Numbers of Classes: Distinct Bone Segmentation in Upper Bodies.}}

%
%
\author{Eva Schnider\inst{1 (}\Envelope\inst{)} \and
Antal Horváth\inst{1}\and
Georg Rauter\inst{1}\and
Azhar Zam\inst{1}\and
Magdalena Müller-Gerbl\inst{2}\and
Philippe C. Cattin\inst{1}}
\authorrunning{E. Schnider et al.}
\titlerunning{3D Networks for Distinct Bone Segmentation in Upper Bodies}
%
\institute{Department of Biomedical Engineering, University of Basel, Allschwil, Switzerland
\email{\{eva.schnider,
	antal.horvath,georg.rauter,azhar.zam,philippe.cattin\}@unibas.ch}\and
Department of Biomedicine, Musculoskeletal Research,
University of Basel, Basel, Switzerland \\
\email{m.mueller-gerbl@unibas.ch}}
\maketitle              
\begin{abstract}
\eva{ 
Segmentation of distinct bones plays a crucial role in diagnosis, planning, navigation, and the assessment of bone metastasis. It supplies semantic knowledge to visualisation tools for the planning of surgical interventions and the education of health professionals. Fully supervised segmentation of 3D data using Deep Learning methods has been extensively studied for many tasks but is usually restricted to distinguishing only a handful of classes. With 125 distinct bones, our case includes many more labels than typical 3D segmentation tasks. For this reason, the direct adaptation of most established methods is not possible. 
This paper discusses the intricacies of training a 3D segmentation network in a many-label setting and shows necessary modifications in network architecture, loss function, and data augmentation. As a result, we demonstrate the robustness of our method by automatically segmenting over one hundred distinct bones simultaneously in an end-to-end learnt fashion from a CT-scan.
}
\keywords{3D segmentation  \and Deep learning \and Many label segmentation}
\end{abstract}
%
%
%
\section{Introduction} 
\eva{
The segmentation of distinct bones from CT images is often performed as an intermediate or preprocessing task for planning and navigation purposes to provide semantic feedback to those systems. It is also crucial for the evaluation of the progress of bone diseases \cite{fu_hierarchical_2017}, or for the quantification of skeletal metastases \cite{lindgren_belal_deep_2019}. 
In Virtual Reality (VR) tools \cite{knodel2018virtual,faludizoller2019haptic}, the distinct segmentation of bones permits more fine-grained control over rendered body parts and can serve an educational purpose by teaching skeletal anatomy.
Due to its distinctive high Hounsfield unit (HU) values in CT images, cortical bone tissue can be segmented approximately using thresholding. However, random intensity variations and the relatively low HU value of squamous bones hinder accurate results \cite{perez2018joint}. For a precise segmentation, or the separation of individual bones, more elaborate methods are needed. For the analysis and segmentation of single bones, statistical shape or appearance models are applied \cite{rahbani2019robust,sarkalkan2014statistical,seim2008automatic}. For whole skeletons, atlas segmentations using articulated joints have been used in mice \cite{baiker2007fullyMice}, and for human upper bodies \cite{fu_hierarchical_2017}. 
A combination of shape models and convolutional neural networks (CNN) have been employed in \cite{lindgren_belal_deep_2019} to segment almost fifty distinct bones. Their multi-step approach consists of an initial shape model corrected landmark detection, followed by a subsequent voxel-wise segmentation. Solely CNN based methods have been used for full-body bone tissue segmentation, without labelling of individual bones \cite{klein2019automatic}, and for segmentation of bones of groups, such as vertebrae \cite{sekuboyina2020verse}. To our knowledge, no simultaneous segmentation of all distinct bones of a human upper body by the use of CNNs has been published so far.

Fully automated methods driven by CNNs have shown great results for various tasks in medical image analysis. They excel at pathology detection \cite{bilic2019LITS,kamnitsas2017efficient,isensee2019nonewnet} as well as at segmenting anatomical structures \cite{horvath2018spinal,lessmann2019iterative,ronneberger2015unet} for a wide array of body regions and in both 2D and 3D. In conjunction with data augmentation, good results have been reported even when training networks on as little as 1-3 fully annotated scans \cite{cicek20163dunet,chaitanya2019semi}.
}
However, in typical 3D medical image segmentation tasks, distinctions are made for a handful or up to a dozen classes. 
Many established methods developed for a few classes fail when dealing with the over hundred classes for our particular case, or are not practical anymore due to restrictions in computational time and memory.

\begin{figure}[b]
\centering
\includegraphics[width=\textwidth]{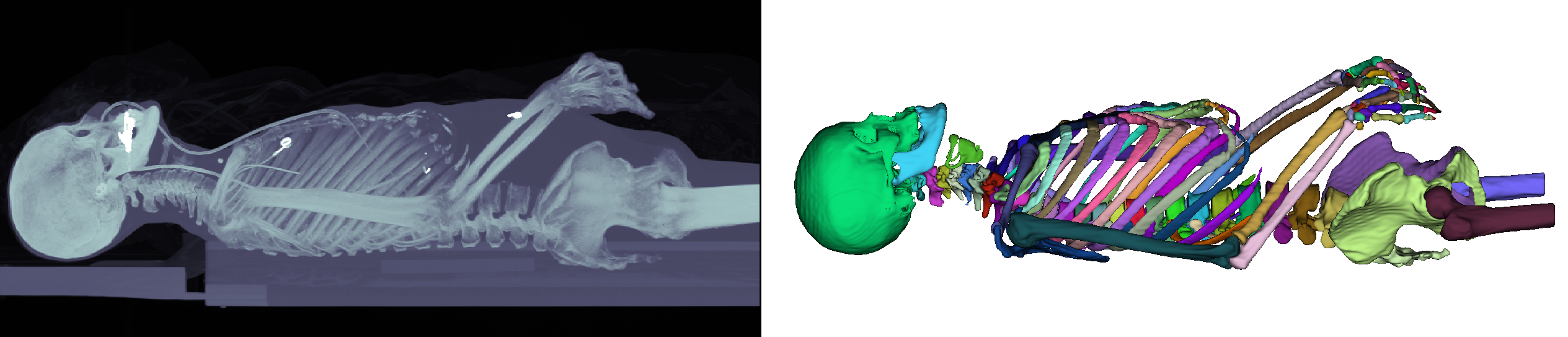}
\caption{Left: Maximum intensity projection of one of our upper body CT-scans. Right: The manual target segmentation depicting 125 different bones with individual colours.} \label{task}
\end{figure}
\eva{
In this work, we present, which kinds of preprocessing, network choice, loss function and data augmentation schemes are suitable for 3D medical image segmentation with many labels at once, using the example of distinct bone segmentation in upper bodies.
Our contributions are: 1) We discuss essential adaptions concerning network choice and data augmentation when performing 3D segmentation in a many-label setting. 2) We examine different sampling strategies and loss functions to mitigate the class imbalance. 3) We present results on a 3D segmentation task with over 100 classes, as depicted in \autoref{task}. }

\section{Methods}
\eva{
Segmenting many classes simultaneously in 3D comes at a cost in computational space and time. In the following, we discuss how this affects and limits not only the possibilities in network design but also renders certain loss functions and data augmentation schemes impractical. We present the methods that worked under the constraints imposed by the many-class task and rendered distinct bone segmentation from upper body CT scans possible. }

\subsection{Limitations imposed by many-label 3D segmentation tasks}
Limitations in computational resources, particularly in GPU RAM size, ask for a careful design of 3D segmentation networks. 
There are many existing architectures optimised for typical GPU memory sizes. 
They generally support input patches in the range of $64^3\,$px to $128^3\,$px and feature only few network layers at the costly full resolution -- mainly input and classification layers. The full resolution classification layer becomes much bigger in the presence of a high number of classes $N_c$, since its size is given by $H\times W \times D \times N_c$, where $H$, $W$, and $D$ represent the output patches' spatial dimensions.

\eva{
One possibility to counter the computational challenges would be splitting of the task into different groups of bones and learning one network per group. Such an ensemble approach has its own downsides, however. There is much overhead needed to train not one, but many networks for the tasks. Apart from training, the added complexity also increases resources and time needed during inference \cite{lee2017overcoming}. Even if resorting to such an approach, both hands alone would sum up to 54 bones (sesamoid bones not included), and therefore considerations about simultaneous segmentation of many bones remain an issue.
}

\subsection{Network design}
For the segmentation task, we use No-New-Net \cite{isensee2019nonewnet}. 
This modification of the standard 3D U-Net \cite{cicek20163dunet} achieves similar performance with less trainable parameters, thus increasing the possible size of input patches and allowing us to capture more global context for our task. 
We were able to use input and output patches of spatial size $96^{3}\,$px on a \SI{8}{\giga \byte}, $128^{3}\,$px on a \SI{12}{\giga \byte}, and of size $160^{3}\,$px on a \SI{24}{\giga \byte} GPU. Even the latter is nowhere near the original size of our CT-scans, the extent of which is \SI{512}{\px} for the smallest dimension. The disparity between scan and patch size means that we can use only a minuscule part of the volume at once and consequently loose information on the global context and surrounding of the subvolume. However, using patches is akin to random cropping of the input and an established technique even for applications where the cropping is not necessary for GPU memory reasons. All in all, we have to balance the increasing information loss of extracting smaller volumes with the enhanced data augmentation effect of more aggressive cropping.

\subsection{Fast balancing many-class segmentation loss}
As a consequence of the unusually large classification layer, any byte additionally spent for representing a single voxel label in the final prediction is amplified millionfold. Using a dense representation of the prediction instead of a sparse one will tip the balance easily towards an out-of-memory error.
We thus use sparse representations of the class-wise predictions and ground truths for computation of the loss. To counter the high imbalance in the number of voxels per class, we use the multi-class cross-entropy loss in conjunction with a Dice similarity coefficient (DSC) loss over all classes $c \in C$: We chose to use an unweighted linear combination of the two, following the implementation given in \cite{isensee2019nonewnet}:

\begin{equation}
    \mathcal{L}_{\text{X-Ent + DSC}} \coloneqq \mathcal{L}_{\text{X-Ent}} + \sum_{c \in C} 
    \mathcal{L}_{\mathrm{DSC}}^{c}.
    \label{xent_plus_dice_loss_def}
\end{equation}

\subsection{Resourceful data augmentation}
\eva{We utilise various data augmentation techniques to increase the variety of data the network encounters during training.} We use random sampling of the input patch locations in two flavours: Uniform random sampling returns every possible patch with the same probability. With balanced sampling, every class has the same probability of being present in the chosen subvolume.
Balanced sampling results in high variability in the field of views of the (input) patches while asserting to repeatedly present all bones, even small ones, to the network.

Much like random cropping, many of the other prevalent techniques in 3D segmentation such as affine transformations, elastic deformations, and changes in brightness and contrast can be employed unhindered in the many-label setting. Contrarily, some augmentation schemes -- notably MixUp \cite{zhang2017mixup} and its variants --  work with dense labels and losses, thus causing tremendous inflation of the classification layer size and loss calculation time. We, therefore, omit the latter kind of data augmentation and concentrate on the first kind.

\subsection{Implementation Details}
Our experiments are built on top of the NiftyNet \cite{gibson2018niftynet} implementation of the No-New-Net \cite{isensee2019nonewnet}. We modified the network architecture only in the number of channels of the classification layer, to account for the different amount of classes. We used the Leaky ReLU activation function with a leak factor of 0.02, and instance normalisation. In contrast to the No-New-Net publication \cite{isensee2019nonewnet}, we were only able to fit a batch size of 1 due to the high memory demands of our many-class case.
We optimised our networks using Adam \cite{kingma2014adam} with a learning rate of 0.001 and ran $20\,000$ iterations of training.

\begin{figure}[t]
\includegraphics[width=\textwidth]{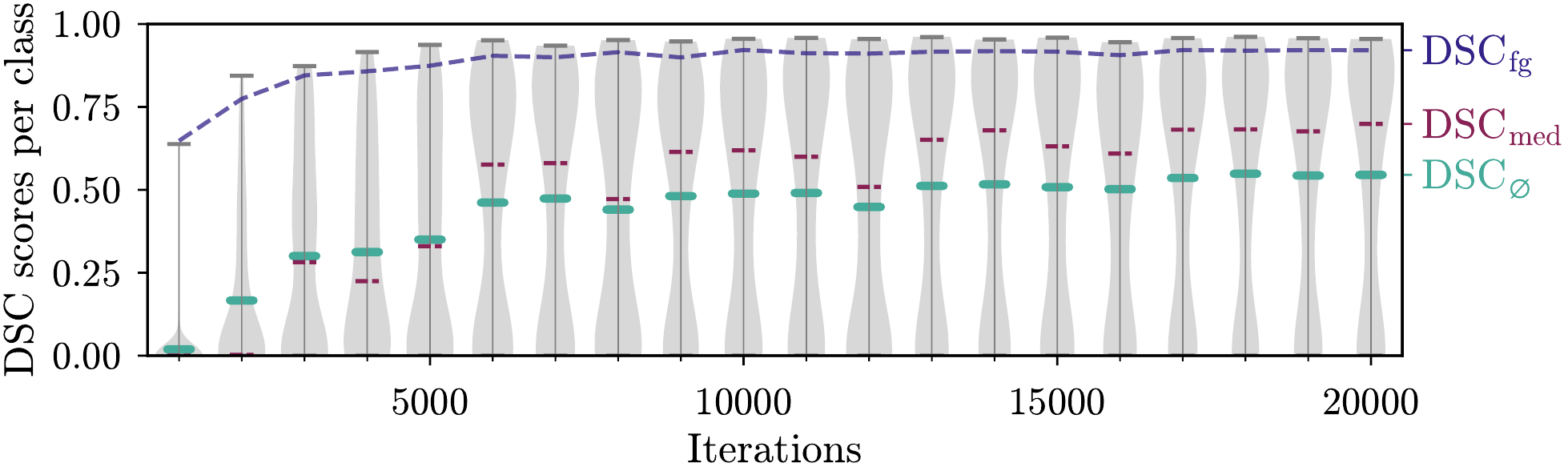}
\caption{Development of DSC scores (see \autoref{metrics}) over the course of training. The distribution of per-class DSC scores is indicated by the violins (grey area). Additionally, the mean, median, and foreground DSC scores \eqref{DSCfg}, are provided.} \label{dice_over_iter}
\end{figure}

\section{Experiments}
For lack of publicly available data sets with many-label distinct bone segmentation, our experiments are conducted on an in-house data set, consisting of five CT scans and their voxel-wise segmentation into 126 classes. To counter the low number of labelled images, we use 5-fold cross-validation throughout.

\subsection{Data Set and Preprocessing}

The five CT scans were taken and annotated by our university's anatomical department. The resulting voxel-wise segmentation consists of 126 classes -- one for each kind of bone in the scans, plus background. The scans were taken from individual subjects aged 44-60, three of whom were female, two male. The field of view starts at the top of the skull and includes the area below until approximately mid-femur. All subjects lie on their backs, arms either resting on the lap or crossed over the stomach, a posture variation that makes the segmentation task harder. The size of each scan was $512 \times 512 \times H$, where the value of $H$ ranges from 656 to 1001. In-plane resolutions vary from $\SI{0.83}{\milli  \metre} \times \SI{0.83}{\milli  \metre}$ to $\SI{0.97}{\milli  \metre} \times \SI{0.97}{\milli  \metre}$ while inter-plane spacing ranges from \SI{1.0}{\milli  \metre} to \SI{1.5}{\milli  \metre}.

\eva{
To be able to capture more body context within an input patch, we resampled our data to \SI{2}{\mm} per dimension -- approximately half the original resolution --  resulting in volumes of }$214-252 \times 215-252 \times 477-514$. We used bilinear interpolation for the scans and nearest neighbour interpolation for the label volume.

\subsection{Evaluation}\label{metrics}

To evaluate the network's ability to correctly label and delineate each bone, we use the DSC of individual classes $c$ in all our experiments: $\mathrm{DSC}_c = \frac{2\abs{P_c\odot G_c}}{\abs{P_c}+\abs{G_c}}$, where $P_c$ and $G_c$ represent the pixel-wise binary form of the prediction of class $c$ and the corresponding ground truth.
\eva{
To obtain a combined score for a whole group of bones over all cross-validation sets, we provide the median DSC. We furthermore provide the distance from the median to the upper and lower uncertainty bound, which correspond to the 16 and 84 percentile. If certain bones are not detected at all, i.e. their DSC equals 0, they are excluded to not distort the distribution. Instead, we provide the detection ratio} 
\begin{equation}
\mathrm{dr}\coloneqq \frac{\#\text{ bones with DSC} > 0}{\#\text{ all bones}}.
\label{detection_ratio}
\end{equation}

Additionally, we provide the foreground (fg) DSC of all bone tissue combined. In this case no distinctions between bones are made. We define the $\mathrm{DSC}_{\mathrm{fg}}$ using foreground ground truth and prediction $G_{\mathrm{fg}}\coloneqq \bigvee_{\substack{c \in C \\ c \neq \mathrm{bg}}} G_{c}$ and  $P_{\mathrm{fg}}\coloneqq \bigvee_{\substack{c \in C \\ c \neq \mathrm{bg}}} P_{c}$.  Assuming mutually exclusive class segmentations we can compute
\begin{equation}
\mathrm{DSC}_{\mathrm{fg}} \coloneqq
\frac{2\abs{P_{\mathrm{fg}}\odot G_{\mathrm{fg}}}}{\abs{P_{\mathrm{fg}}}+\abs{G_{\mathrm{fg}}}}
= \frac{2\abs{\overline{P_\mathrm{bg}}\odot \overline{G_\mathrm{bg}}}}
{\abs{\overline{P_\mathrm{bg}}}+\abs{\overline{G_\mathrm{bg}}}},
\label{DSCfg}
\end{equation}

 using only the background segmentation. In this equation, $ \overline{P_\mathrm{bg}}$ denotes the logic complement of the binary predication for the background class bg, and $\overline{G_\mathrm{bg}}$ denotes the respective ground truth. 

We employ cross-validation using five different data folds, each comprising of three scans for training, one for validation and one for testing. 
The validation set is used for adjusting the hyperparameters and monitoring convergence. 
Within every cross-evaluation fold, we use a different scan for the final testing.

\section{Results and Discussion}

\begin{table}[t]
\caption{
\eva{Comparison of segmentation performance per model. 
We provide the median DSC, the uncertainty boundaries, along with the detection ratio dr \eqref{detection_ratio} for each group of bones, and the median foreground DSC \eqref{DSCfg} for all bones combined. Time per training iteration normalised by batch size.} 
}\label{table_model_compare}
\begin{tabular}{l c l c c l c c l c c l c c l c c c r}
\toprule
\multicolumn{1}{l}{Method}    && \multicolumn{15}{c}{Segmentation performance for groups of bones} && Time\\

\cmidrule{3-17}

&& \multicolumn{3}{c}{Spine} & \multicolumn{3}{c}{Ribs} & \multicolumn{3}{c}{Hands} & \multicolumn{3}{c}{Large bones} & \multicolumn{4}{c}{All}&\\


&&  \multicolumn{1}{c}{DSC}  & \multicolumn{1}{c}{dr} & \phantom{i}& \multicolumn{1}{c}{DSC} & \multicolumn{1}{c}{dr} &\phantom{i}& \multicolumn{1}{c}{DSC}    & \multicolumn{1}{c}{dr} &\phantom{i}&\multicolumn{1}{c}{DSC}    &\multicolumn{1}{c}{dr}&\phantom{i}& \multicolumn{1}{c}{DSC}   & \multicolumn{1}{c}{dr} & \multicolumn{1}{c}{fg} &\phantom{i}&\multicolumn{1}{r}{s}\\
\midrule

$\mathrm{96_{bal}}$&& $ 0.79_{-0.26}^{+0.11}$ & 1 && $ 0.52_{-0.26}^{+0.26}$ & 1 && $ 0.48_{-0.38}^{+0.31}$ & 0.54 && $ 0.83_{-0.19}^{+0.07}$ & 1 && $ 0.68_{-0.43}^{+0.19}$ & 0.79 & 0.84&& 2.1 \\ 
\addlinespace
$\mathrm{96_{bal,xent}  }$&& $ 0.81_{-0.39}^{+0.09}$ & 1 && $ 0.53_{-0.27}^{+0.21}$ & 1 && $ 0.42_{-0.35}^{+0.37}$ & 0.57 && $ 0.87_{-0.09}^{+0.05}$ & 1 && $ 0.66_{-0.40}^{+0.21}$ & 0.80 & 0.90&& 1.1 \\ 
\addlinespace
$\mathrm{128_{unif,d}  }$&& $ 0.80_{-0.20}^{+0.09}$ & 1 && $ 0.62_{-0.32}^{+0.20}$ & 1 && $ 0.52_{-0.42}^{+0.21}$ & 0.41 && $ 0.90_{-0.04}^{+0.04}$ & 1 && $ 0.73_{-0.38}^{+0.16}$ & 0.73 & 0.89&& 5.2 \\ 
\addlinespace
$\mathrm{128_{bal,d}  }$&& $ 0.80_{-0.28}^{+0.11}$ & 1 && $ 0.54_{-0.35}^{+0.23}$ & 1 && $ 0.58_{-0.46}^{+0.27}$ & 0.51 && $ 0.84_{-0.17}^{+0.07}$ & 1 && $ 0.71_{-0.48}^{+0.18}$ & 0.77 & 0.85&& 5.3 \\ 
\addlinespace
$\mathrm{160_{bal,d}}$&& $ 0.82_{-0.17}^{+0.09}$ & 1 && $ 0.58_{-0.27}^{+0.21}$ & 1 && $ 0.67_{-0.39}^{+0.18}$ & 0.58 && $ 0.88_{-0.11}^{+0.04}$ & 1 && $ 0.75_{-0.38}^{+0.14}$ & 0.80 & 0.88&& 8.8 \\ 
\addlinespace
$\mathrm{160_{bal,xent,d}}$&& $ 0.83_{-0.25}^{+0.09}$ & 1 && $ 0.58_{-0.29}^{+0.23}$ & 1 && $ 0.55_{-0.41}^{+0.28}$ & 0.59 && $ 0.90_{-0.08}^{+0.04}$ & 1 && $ 0.75_{-0.43}^{+0.15}$ & 0.81 & 0.89&& 3.7 \\ 
\addlinespace

\midrule
$\text{2D U-Net}_{2c}$  && \multicolumn{1}{c}{--}&\multicolumn{1}{c}{--}&& \multicolumn{1}{c}{--}& \multicolumn{1}{c}{--}&& \multicolumn{1}{c}{--}& \multicolumn{1}{c}{--}&& \multicolumn{1}{c}{--}& \multicolumn{1}{c}{--}&& \multicolumn{1}{c}{--}& \multicolumn{1}{c}{--}& 0.91&& 0.4\\
$\text{2D U-Net}_{126c}$&& $ 0.45_{-0.30}^{+0.24}$ & 0.87 && $ 0.34_{-0.27}^{+0.26}$ & 0.94 && $ 0.36_{-0.26}^{+0.33}$ & 0.23 && $ 0.82_{-0.19}^{+0.08}$ & 1 && $ 0.49_{-0.37}^{+0.29}$ & 0.61 & 0.86&& 0.4 \\ 
\addlinespace

\bottomrule
\end{tabular}
\end{table}

\eva{
To evaluate the contributions of different patch sizes, sampling strategies, data augmentation schemes and loss functions, we present quantitative results in \autoref{table_model_compare}.} 
We investigate input patch sizes of 96, 128, and 160\,px per dimension, chosen through balanced sampling {\small\textbf{bal}} or uniform sampling {\small\textbf{unif}}. The subscript {\small\textbf{xent}} stands for the use of the cross-entropy loss function alone instead of the full loss  \eqref{xent_plus_dice_loss_def}. With {\small\textbf{d}} we denote data augmentation with elastic deformations.

\eva{Not least because of the small data set available, there is considerable variance within the DSC scores of a given model and bone group, which impedes direct comparison of different models.
No single model outperforms all others, although bigger patch sizes correspond to higher total scores. As for class imbalances, we note that the two models trained with a uniform sampler have the lowest detection ratio for bones in the hands. The balanced sampler thus seems to benefit the detection and segmentation of tiny bones.
We indicate the time needed for one iteration of training.
To ensure a fair comparison, we averaged 100 iterations trained on the same machine under stable conditions.
Patch sizes profoundly influence the time needed per iteration. 
The resulting times range from close to 1 second for a patch of size $96^3$ up to almost 9 seconds for patches sized $160^3$. 
The loss function also influences the training time considerably, with pure cross-entropy taking only half as long as the combined loss function.
}

\eva{Because many of our limitations in computational resources stem from the combination of a 3D network with a large number of classes, we additionally provide the results obtained using a 2D U-Net. 
We trained this network as specified in \cite{klein2019automatic} who used it successfully for non-distinct bone segmentation of whole-body CT images. 
This network leads to good results for the 2-class case ($\text{2D U-Net}_{2c}$), but it does not scale well to bone distinction, as our results of a $\text{2D U-Net}_{126c}$ -- trained on the primary task -- suggest.}

\begin{figure}[t]
\includegraphics[width=\textwidth]{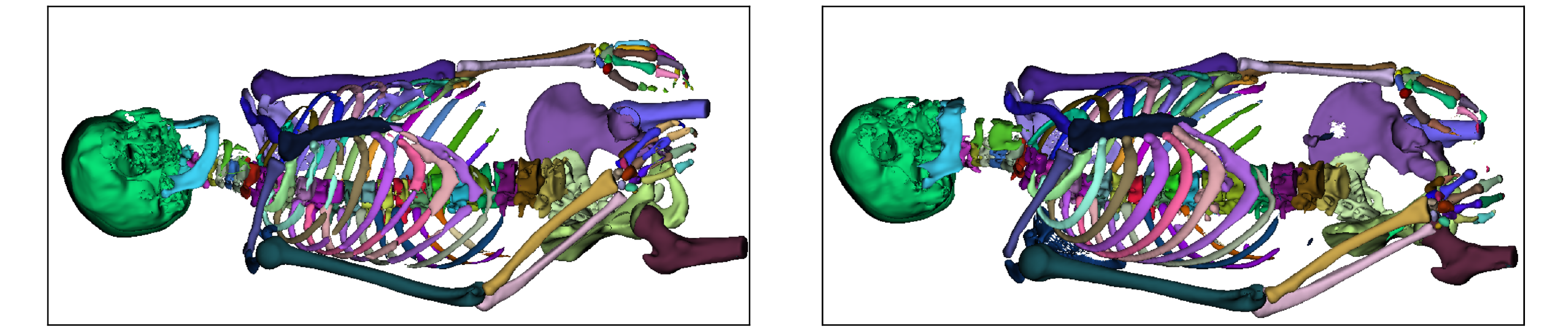}
\caption{\eva{Qualitative segmentation results created using $\mathrm{160_{bal,d}}$ and two exemplary CT scans for which no manual labels exist. The 3D views were created with 3D Slicer \cite{slicer20123d} and show an interpolated version of the data.}}
\label{results}
\end{figure}

\eva{A comparison with existing methods is made in \autoref{comparison_others}. 
Since code and data sets are not publicly available, we compare the published results for different bones with our own. 
While the atlas method presented in \cite{fu_hierarchical_2017} exhibits the best segmentation performance, their inference takes 20 minutes. They also require manual intervention if used on CT images that show only parts of an upper body. The two-step neural network presented in \cite{lindgren_belal_deep_2019} was trained on 100 data sets and evaluated on 5. For the sacrum and L3, both our work and \cite{lindgren_belal_deep_2019} show similar results. For bones that have a high chance of being confused with the ones below and above, their use of a shape model for landmark labelling and post-processing helps to keep scores for ribs and vertebrae high. It is, however, not clear how easily their approach could be adapted to accommodate for the segmentation of further bones, e.g. hands.

Qualitative results using two scans of unlabelled data are depicted in \autoref{results}.
}

\begin{table}[t]
\caption{\eva{Comparison of segmentation results for an end-to-end trained neural network approach (this work, model $\mathrm{160_{bal}}$), a hybrid approach using neural networks and shape models for landmark detection and a subsequent neural network for segmentation \cite{lindgren_belal_deep_2019}, and a hierarchical atlas segmentation \cite{fu_hierarchical_2017}.  } }\label{comparison_others}
\begin{tabular}{l c l l c l l c l l c }

\toprule
\multicolumn{1}{l}{\phantom{i}}&& \multicolumn{2}{c}{this work} && \multicolumn{2}{c}{Lindgren et al. \cite{lindgren_belal_deep_2019}} && \multicolumn{2}{c}{Fu et al. \cite{fu_hierarchical_2017}}& \\
\cmidrule{3-4}\cmidrule{6-7}\cmidrule{9-10}


\multicolumn{1}{l}{DSC}&&  \multicolumn{1}{l}{median \phantom{i}}  & \multicolumn{1}{l}{range} & \phantom{M}& \multicolumn{1}{l}{median \phantom{i}} & \multicolumn{1}{l}{range} &\phantom{M}& \multicolumn{1}{l}{$\varnothing_c$ \phantom{iiii}}    & \multicolumn{1}{l}{std} &\phantom{i} \\
\midrule

Th7&& 0.64 & 0.22-0.94 && 0.86 & 0.42-0.89&& 0.85 & 0.02&\\
L3   && 0.89 & 0.72-0.94 && 0.85 & 0.72-0.90&& 0.91 & 0.01&\\
Sacrum   && 0.86 & 0.80-0.92 && 0.88 & 0.76-0.89&& -- & --& \\
Rib  && 0.38 & 0.19-0.58 && 0.84 & 0.76-0.86&& -- & --&\\
Sternum && 0.74 & 0.59-0.87&& 0.83 & 0.80-0.87&& 0.89 & 0.02 &\\
\midrule
Inference time for 1 scan (min) && \multicolumn{2}{c}{$\sim1$\phantom{$\sim$}} &&  \multicolumn{2}{c}{--} && \multicolumn{2}{c}{$\sim20$\phantom{$\sim$}} &\\
Distinct bones (\#)&& \multicolumn{2}{c}{125} &&  \multicolumn{2}{c}{49} && \multicolumn{2}{c}{62} &\\
In-plane resolution (mm) && \multicolumn{2}{c}{2} &&  \multicolumn{2}{c}{3.27} &&\multicolumn{2}{c}{0.97} &\\
Slice thickness (mm) && \multicolumn{2}{c}{2} &&  \multicolumn{2}{c}{3.75} &&\multicolumn{2}{c}{1.5-2.5} &\\
\bottomrule
\end{tabular}
\end{table}
\section{Summary and Conclusion}
\eva{
We tackled the task of segmenting 125 distinct bones at once in an upper-body CT scan, using an end-to-end trained neural network and only three fully labelled scans for training. We provide network architectures, loss functions and data augmentation schemes which make this computationally singular task feasible. While not all problems are solved, we showed how balanced sampling and a suitable choice of the loss function help to deal with the class imbalance inherent to our task.
Despite a lack of training data, we obtained median DSC scores of up to 0.9 on large bones, 0.8 on vertebrae, which compares well with other works that segment various bones of the upper body simultaneously. More problematic are ribs, which tend to be confused with one another, an issue where shape models certainly could help. As for the hands, many of the tiny bones are not detected at all, which suggests the need for more fine-grained methods for this particular set of bones. In terms of inference time, the complete labelling of a scan takes roughly one minute, which would be fast enough to be used to create initial guesses of a more accurate atlas method.
More manually labelled scans would certainly improve the generalisation capacity of our networks and the statistical significance of our comparisons. Using our results on momentarily unlabelled data as priors, we expect a drastic decrease in the time needed for further manual annotations.
}

\subsubsection{Acknowledgements}
This work was financially supported by the Werner Siemens
Foundation through the MIRACLE project. We thank Mireille Toranelli for acquiring the scans and providing the ground truth labelling.

%


\bibliographystyle{splncs04}
\bibliography{bibliography}

\end{document}


\title{Many Labels, Little Data -- Automated 3D Segmentation of Distinct Bones in Upper Bodies.\thanks{Supported by Anonymous Organisation.}}

%
%
\author{Anonymous\inst{1} \and
Anonymous\inst{1}\and
Anonymous\inst{1}\and
Anonymous\inst{1}\and
Anonymous\inst{1, 2}\and
Anonymous\inst{1}}
\authorrunning{A. Nonymous et al.}
\titlerunning{Many Labels, Little Data -- 3D Segmentation of Distinct Bones}
%
\institute{Anonymous Institute, University, Town, Country
\email{\{********.*******,**********.*********,*******.*******\}@*******.**}\and
Anonymous Institute, University, Town, Country \\
\email{*******.*******@*******.**}}
%

%
%
%
\section*{Supplementary Material} 

\subsection*{3D Reconstructions of All Available Test Data}
\begin{figure}[h!]
\centering
\includegraphics[width=\textwidth]{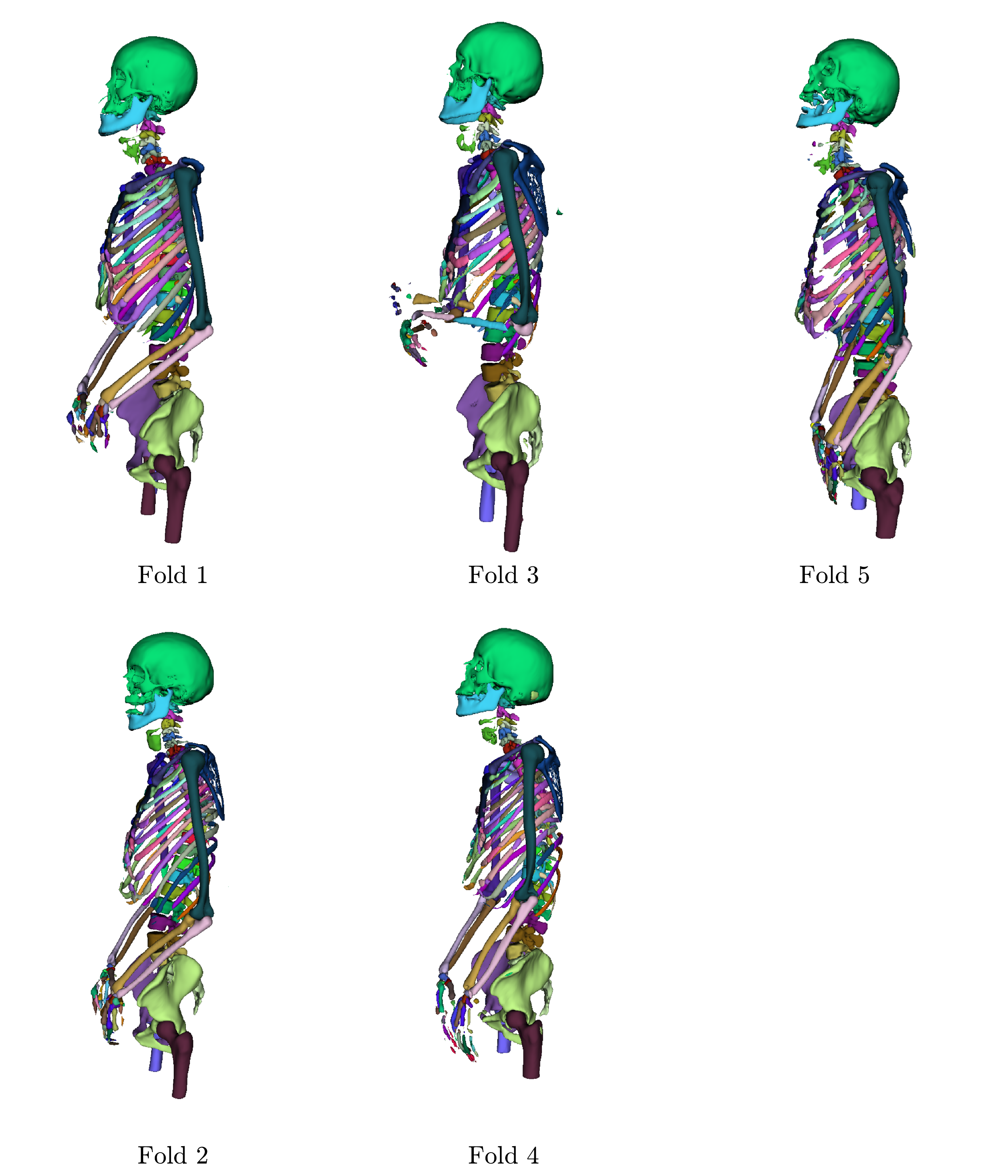}
\caption{Predictions created with $\mathrm{Ba_{128,un}}$ on the test set of every cross-validation fold. The interpolated 3D reconstructions were made using 3D slicer. Different colours represent different segmentation classes as listed in \autoref{tab1}.} \label{3dplots}
\end{figure}

\definecolor{0}{RGB}{255, 255, 255}
\definecolor{1}{RGB}{126, 11, 152}
\definecolor{2}{RGB}{89, 159, 106}
\definecolor{3}{RGB}{44, 229, 40}
\definecolor{4}{RGB}{40, 185, 210}
\definecolor{5}{RGB}{213, 52, 194}
\definecolor{6}{RGB}{133, 166, 116}
\definecolor{7}{RGB}{169, 114, 181}
\definecolor{8}{RGB}{167, 24, 189}
\definecolor{9}{RGB}{57, 118, 100}
\definecolor{10}{RGB}{103, 165, 170}
\definecolor{11}{RGB}{196, 220, 72}
\definecolor{12}{RGB}{224, 25, 59}
\definecolor{13}{RGB}{118, 155, 121}
\definecolor{14}{RGB}{162, 51, 7}
\definecolor{15}{RGB}{37, 20, 198}
\definecolor{16}{RGB}{79, 80, 156}
\definecolor{17}{RGB}{61, 178, 235}
\definecolor{18}{RGB}{100, 22, 192}
\definecolor{19}{RGB}{123, 148, 95}
\definecolor{20}{RGB}{72, 39, 146}
\definecolor{21}{RGB}{29, 85, 93}
\definecolor{22}{RGB}{46, 201, 201}
\definecolor{23}{RGB}{209, 63, 212}
\definecolor{24}{RGB}{174, 188, 48}
\definecolor{25}{RGB}{201, 223, 188}
\definecolor{26}{RGB}{64, 125, 217}
\definecolor{27}{RGB}{158, 193, 162}
\definecolor{28}{RGB}{178, 35, 19}
\definecolor{29}{RGB}{85, 196, 47}
\definecolor{30}{RGB}{242, 250, 5}
\definecolor{31}{RGB}{160, 153, 214}
\definecolor{32}{RGB}{65, 212, 246}
\definecolor{33}{RGB}{94, 76, 75}
\definecolor{34}{RGB}{108, 70, 8}
\definecolor{35}{RGB}{15, 194, 131}
\definecolor{36}{RGB}{73, 64, 250}
\definecolor{37}{RGB}{121, 77, 64}
\definecolor{38}{RGB}{148, 112, 90}
\definecolor{39}{RGB}{222, 174, 15}
\definecolor{40}{RGB}{81, 21, 251}
\definecolor{41}{RGB}{101, 209, 245}
\definecolor{42}{RGB}{187, 197, 98}
\definecolor{43}{RGB}{218, 247, 173}
\definecolor{44}{RGB}{188, 152, 27}
\definecolor{45}{RGB}{208, 186, 234}
\definecolor{46}{RGB}{203, 167, 77}
\definecolor{47}{RGB}{7, 230, 179}
\definecolor{48}{RGB}{165, 251, 163}
\definecolor{49}{RGB}{88, 142, 213}
\definecolor{50}{RGB}{153, 12, 225}
\definecolor{51}{RGB}{68, 163, 14}
\definecolor{52}{RGB}{207, 95, 4}
\definecolor{53}{RGB}{109, 33, 247}
\definecolor{54}{RGB}{221, 207, 161}
\definecolor{55}{RGB}{98, 82, 11}
\definecolor{56}{RGB}{155, 253, 229}
\definecolor{57}{RGB}{22, 32, 244}
\definecolor{58}{RGB}{141, 126, 88}
\definecolor{59}{RGB}{214, 147, 186}
\definecolor{60}{RGB}{192, 93, 243}
\definecolor{61}{RGB}{134, 111, 239}
\definecolor{62}{RGB}{252, 103, 166}
\definecolor{63}{RGB}{42, 213, 185}
\definecolor{64}{RGB}{227, 170, 177}
\definecolor{65}{RGB}{145, 200, 114}
\definecolor{66}{RGB}{175, 106, 228}
\definecolor{67}{RGB}{233, 160, 49}
\definecolor{68}{RGB}{166, 183, 157}
\definecolor{69}{RGB}{181, 0, 202}
\definecolor{70}{RGB}{19, 74, 130}
\definecolor{71}{RGB}{70, 94, 200}
\definecolor{72}{RGB}{193, 173, 149}
\definecolor{73}{RGB}{149, 133, 231}
\definecolor{74}{RGB}{21, 72, 148}
\definecolor{75}{RGB}{8, 225, 119}
\definecolor{76}{RGB}{17, 30, 62}
\definecolor{77}{RGB}{164, 8, 1}
\definecolor{78}{RGB}{25, 2, 232}
\definecolor{79}{RGB}{39, 243, 96}
\definecolor{80}{RGB}{190, 43, 107}
\definecolor{81}{RGB}{114, 161, 21}
\definecolor{82}{RGB}{62, 198, 151}
\definecolor{83}{RGB}{104, 205, 29}
\definecolor{84}{RGB}{216, 128, 23}
\definecolor{85}{RGB}{74, 169, 70}
\definecolor{86}{RGB}{173, 109, 87}
\definecolor{87}{RGB}{140, 192, 2}
\definecolor{88}{RGB}{119, 55, 43}
\definecolor{89}{RGB}{226, 36, 112}
\definecolor{90}{RGB}{11, 9, 209}
\definecolor{91}{RGB}{246, 141, 199}
\definecolor{92}{RGB}{238, 195, 123}
\definecolor{93}{RGB}{52, 92, 55}
\definecolor{94}{RGB}{77, 86, 65}
\definecolor{95}{RGB}{27, 156, 86}
\definecolor{96}{RGB}{75, 57, 38}
\definecolor{97}{RGB}{107, 90, 129}
\definecolor{98}{RGB}{135, 175, 143}
\definecolor{99}{RGB}{217, 204, 3}
\definecolor{100}{RGB}{38, 18, 10}
\definecolor{101}{RGB}{91, 5, 84}
\definecolor{102}{RGB}{247, 19, 79}
\definecolor{103}{RGB}{151, 13, 191}
\definecolor{104}{RGB}{9, 105, 67}
\definecolor{105}{RGB}{146, 59, 187}
\definecolor{106}{RGB}{200, 50, 53}
\definecolor{107}{RGB}{106, 190, 63}
\definecolor{108}{RGB}{212, 249, 171}
\definecolor{109}{RGB}{182, 28, 76}
\definecolor{110}{RGB}{90, 208, 204}
\definecolor{111}{RGB}{139, 214, 73}
\definecolor{112}{RGB}{63, 66, 154}
\definecolor{113}{RGB}{136, 99, 54}
\definecolor{114}{RGB}{232, 191, 219}
\definecolor{115}{RGB}{131, 87, 172}
\definecolor{116}{RGB}{194, 248, 137}
\definecolor{117}{RGB}{115, 104, 248}
\definecolor{118}{RGB}{95, 42, 66}
\definecolor{119}{RGB}{142, 176, 33}
\definecolor{120}{RGB}{6, 168, 94}
\definecolor{121}{RGB}{124, 23, 128}
\definecolor{122}{RGB}{127, 91, 20}
\definecolor{123}{RGB}{189, 179, 83}
\definecolor{124}{RGB}{32, 81, 30}
\definecolor{125}{RGB}{205, 244, 176}

\begin{table}[b]
\caption{All classes used for distinct bone segmentation of upper bodies. Additionally, the colour coding that has been applied to plots is provided. The two sides of the body are abbreviated with le and ri for left and right, respectively.}\label{tab1}
\begin{tabular}{r c l c l r c l c l r c l c l}
\toprule
No. &\phantom{i}& Name & Colour &\phantom{M}& No. &\phantom{i}& Name & Colour &\phantom{M}& No. &\phantom{i}& Name & Colour \\

\midrule

0 && Background & \fcolorbox{black}{0}{\phantom{m}}&&42 && Metacarpale V le & \colorbox{42}{\phantom{M}}&&84 && Phalanx I dist. le & \colorbox{84}{\phantom{M}}\\
1 && Th1 & \colorbox{1}{\phantom{M}}&&43 && Os pisiforme ri  & \colorbox{43}{\phantom{M}}&&85 && Phalanx I prox. ri  & \colorbox{85}{\phantom{M}}\\
2 && Th10 & \colorbox{2}{\phantom{M}}&&44 && Os pisiforme le & \colorbox{44}{\phantom{M}}&&86 && Phalanx I prox. le & \colorbox{86}{\phantom{M}}\\
3 && Th11 & \colorbox{3}{\phantom{M}}&&45 && Radius ri  & \colorbox{45}{\phantom{M}}&&87 && Phalanx II dist. ri  & \colorbox{87}{\phantom{M}}\\
4 && Th12 & \colorbox{4}{\phantom{M}}&&46 && Radius le & \colorbox{46}{\phantom{M}}&&88 && Phalanx II dist. le & \colorbox{88}{\phantom{M}}\\
5 && Th2 & \colorbox{5}{\phantom{M}}&&47 && Rib 1 ri  & \colorbox{47}{\phantom{M}}&&89 && Phalanx II med. ri  & \colorbox{89}{\phantom{M}}\\
6 && Th3 & \colorbox{6}{\phantom{M}}&&48 && Rib 1 le & \colorbox{48}{\phantom{M}}&&90 && Phalanx II med. le & \colorbox{90}{\phantom{M}}\\
7 && Th4 & \colorbox{7}{\phantom{M}}&&49 && Rib 10 ri  & \colorbox{49}{\phantom{M}}&&91 && Phalanx II prox. ri  & \colorbox{91}{\phantom{M}}\\
8 && Th5 & \colorbox{8}{\phantom{M}}&&50 && Rib 10 le & \colorbox{50}{\phantom{M}}&&92 && Phalanx II prox. le & \colorbox{92}{\phantom{M}}\\
9 && Th6 & \colorbox{9}{\phantom{M}}&&51 && Rib 11 ri  & \colorbox{51}{\phantom{M}}&&93 && Phalanx III dist. ri  & \colorbox{93}{\phantom{M}}\\
10 && Th7 & \colorbox{10}{\phantom{M}}&&52 && Rib 11 le & \colorbox{52}{\phantom{M}}&&94 && Phalanx III dist. le & \colorbox{94}{\phantom{M}}\\
11 && Th8 & \colorbox{11}{\phantom{M}}&&53 && Rib 12 ri  & \colorbox{53}{\phantom{M}}&&95 && Phalanx III med. ri  & \colorbox{95}{\phantom{M}}\\
12 && Th9 & \colorbox{12}{\phantom{M}}&&54 && Rib 12 le & \colorbox{54}{\phantom{M}}&&96 && Phalanx III med. le & \colorbox{96}{\phantom{M}}\\
13 && Os capitatum ri  & \colorbox{13}{\phantom{M}}&&55 && Rib 2 ri  & \colorbox{55}{\phantom{M}}&&97 && Phalanx III prox. ri  & \colorbox{97}{\phantom{M}}\\
14 && Os capitatum le & \colorbox{14}{\phantom{M}}&&56 && Rib 2 le & \colorbox{56}{\phantom{M}}&&98 && Phalanx III prox. le & \colorbox{98}{\phantom{M}}\\
15 && Clavicula ri  & \colorbox{15}{\phantom{M}}&&57 && Rib 3 ri  & \colorbox{57}{\phantom{M}}&&99 && Phalanx IV dist. ri  & \colorbox{99}{\phantom{M}}\\
16 && Clavicula le & \colorbox{16}{\phantom{M}}&&58 && Rib 3 le & \colorbox{58}{\phantom{M}}&&100 && Phalanx IV dist. le & \colorbox{100}{\phantom{M}}\\
17 && Os hamatum ri  & \colorbox{17}{\phantom{M}}&&59 && Rib 4 ri  & \colorbox{59}{\phantom{M}}&&101 && Phalanx IV med. ri  & \colorbox{101}{\phantom{M}}\\
18 && Os hamatum le & \colorbox{18}{\phantom{M}}&&60 && Rib 4 le & \colorbox{60}{\phantom{M}}&&102 && Phalanx IV med. le & \colorbox{102}{\phantom{M}}\\
19 && Pace maker & \colorbox{19}{\phantom{M}}&&61 && Rib 5 ri  & \colorbox{61}{\phantom{M}}&&103 && Phalanx IV prox. ri  & \colorbox{103}{\phantom{M}}\\
20 && Humerus ri  & \colorbox{20}{\phantom{M}}&&62 && Rib 5 le & \colorbox{62}{\phantom{M}}&&104 && Phalanx IV prox. le & \colorbox{104}{\phantom{M}}\\
21 && Humerus le & \colorbox{21}{\phantom{M}}&&63 && Rib 6 ri  & \colorbox{63}{\phantom{M}}&&105 && Phalanx V dist. ri  & \colorbox{105}{\phantom{M}}\\
22 && C1 & \colorbox{22}{\phantom{M}}&&64 && Rib 6 le & \colorbox{64}{\phantom{M}}&&106 && Phalanx V dist. le & \colorbox{106}{\phantom{M}}\\
23 && C2 & \colorbox{23}{\phantom{M}}&&65 && Rib 7 ri  & \colorbox{65}{\phantom{M}}&&107 && Phalanx V med. ri  & \colorbox{107}{\phantom{M}}\\
24 && C3 & \colorbox{24}{\phantom{M}}&&66 && Rib 7 le & \colorbox{66}{\phantom{M}}&&108 && Phalanx V med. le & \colorbox{108}{\phantom{M}}\\
25 && C4 & \colorbox{25}{\phantom{M}}&&67 && Rib 8 ri  & \colorbox{67}{\phantom{M}}&&109 && Phalanx V prox. ri  & \colorbox{109}{\phantom{M}}\\
26 && C5 & \colorbox{26}{\phantom{M}}&&68 && Rib 8 le & \colorbox{68}{\phantom{M}}&&110 && Phalanx V prox. le & \colorbox{110}{\phantom{M}}\\
27 && C6 & \colorbox{27}{\phantom{M}}&&69 && Rib 9 ri  & \colorbox{69}{\phantom{M}}&&111 && Ossa sesamoidea ri  & \colorbox{111}{\phantom{M}}\\
28 && C7 & \colorbox{28}{\phantom{M}}&&70 && Rib 9 le & \colorbox{70}{\phantom{M}}&&112 && Ossa sesamoidea le & \colorbox{112}{\phantom{M}}\\
29 && Os hyoideum & \colorbox{29}{\phantom{M}}&&71 && Os scaphoideum ri  & \colorbox{71}{\phantom{M}}&&113 && Ulna ri  & \colorbox{113}{\phantom{M}}\\
30 && Os lunatum ri  & \colorbox{30}{\phantom{M}}&&72 && Os scaphoideum le & \colorbox{72}{\phantom{M}}&&114 && Ulna le & \colorbox{114}{\phantom{M}}\\
31 && Os lunatum le & \colorbox{31}{\phantom{M}}&&73 && Scapula ri  & \colorbox{73}{\phantom{M}}&&115 && Os coxae ri  & \colorbox{115}{\phantom{M}}\\
32 && Mandibula & \colorbox{32}{\phantom{M}}&&74 && Scapula le & \colorbox{74}{\phantom{M}}&&116 && Os coxae le & \colorbox{116}{\phantom{M}}\\
33 && Metacarpale I ri  & \colorbox{33}{\phantom{M}}&&75 && Cranuim & \colorbox{75}{\phantom{M}}&&117 && Femur ri  & \colorbox{117}{\phantom{M}}\\
34 && Metacarpale I le & \colorbox{34}{\phantom{M}}&&76 && Sternum & \colorbox{76}{\phantom{M}}&&118 && Femur le & \colorbox{118}{\phantom{M}}\\
35 && Metacarpale II ri  & \colorbox{35}{\phantom{M}}&&77 && Os trapezium ri  & \colorbox{77}{\phantom{M}}&&119 && L1 & \colorbox{119}{\phantom{M}}\\
36 && Metacarpale II le & \colorbox{36}{\phantom{M}}&&78 && Os trapezium le & \colorbox{78}{\phantom{M}}&&120 && L2 & \colorbox{120}{\phantom{M}}\\
37 && Metacarpale III ri  & \colorbox{37}{\phantom{M}}&&79 && Os trapezoideum ri  & \colorbox{79}{\phantom{M}}&&121 && L3 & \colorbox{121}{\phantom{M}}\\
38 && Metacarpale III le & \colorbox{38}{\phantom{M}}&&80 && Os trapezoideum le & \colorbox{80}{\phantom{M}}&&122 && L4 & \colorbox{122}{\phantom{M}}\\
39 && Metacarpale IV ri  & \colorbox{39}{\phantom{M}}&&81 && Os triquetrum ri  & \colorbox{81}{\phantom{M}}&&123 && L5 & \colorbox{123}{\phantom{M}}\\
40 && Metacarpale IV le & \colorbox{40}{\phantom{M}}&&82 && Os triquetrum le & \colorbox{82}{\phantom{M}}&&124 && L6 & \colorbox{124}{\phantom{M}}\\
41 && Metacarpale V ri  & \colorbox{41}{\phantom{M}}&&83 && Phalanx I dist. ri  & \colorbox{83}{\phantom{M}}&&125 && Sacrum & \colorbox{125}{\phantom{M}}\\

\bottomrule
\end{tabular}
\end{table}